\documentclass{webofc}

\usepackage[varg]{txfonts}   
\usepackage{hyperref}
\usepackage{url}
\hypersetup{colorlinks=true,citecolor=blue,urlcolor=blue,linkcolor=blue}
\begin{document}
\title{Unconventional Searches for Exotic Particles at Future Lepton Colliders}
%
%

\author{\firstname{Nilanjana} \lastname{Kumar}\inst{1}\fnsep\thanks{\email{nilanjana.kumar@gmail.com}} 
}

\institute{ Centre for Cosmology and Science Popularization, SGT University, Gurugram (Haryana)-122505, India
          }

\abstract{The main aim of the the Large Hadron Collider (LHC) experiments is to search for exotic particles with masses in the TeV range as predicted by Beyond Standard Model (BSM) theories. 
However, there is no hint of BSM around TeV scale so far. Hence, it is possible that the exotic particles are heavier and larger centre of mass energy is needed to observe them. Alternatively, the future lepton colliders offer a comparatively cleaner environment than the LHC which is advantageous to detect light exotic particles. Lepton colliders, like the International Linear Collider, provide the opportunity to detect exotic particles at energies below the TeV scale. The Muon Collider, once fully operational, will have the capability to observe exotic particles at and beyond the TeV scale.
The search for BSM particles 
typically assumes a minimal scenario where only one type of BSM particle couples with the Standard Model (SM) sector. 
But there are theories which involve such interactions of multiple BSM particles. Here I discusses a specific model featuring a 
fermionic quintuplet and a scalar quartet that interact before decaying into SM particles. 
This model yields distinctive signatures characterized by high lepton and jet multiplicities, 
making it a promising candidate for detection at future lepton colliders.
}
\maketitle
%
\section{Introduction}
\label{intro}

Currently the experiments at the Large Hadron Collider are actively searching for elusive exotic particles. 
LHC is currently operating at 13.6 GeV and no BSM particles with mass around TeV scale have been observed so far \cite{CMS:2024bni}. 
To increase the search reach one can either increase the centre of mass energy of the LHC or propose colliders with 
much larger energies\cite{FCC:2018vvp}.
An alternative way is to search for the exotic particles at the future lepton colliders, which 
offer much cleaner environment than the hadron colliders.
Future lepton colliders will start operating on much smaller energies than LHC. For example, 
International Linear Collider \cite{ILCInternationalDevelopmentTeam:2022izu,Zarnecki:2020ics} will operate on 
250 GeV, 500 GeV and 1 TeV. The primary objectives of the ILC experiments are precession measurements and 
properties of the Higgs boson\cite{Tian:2016qlk}. Additionally, ILC will serve as an exceptional machine for detecting BSM 
particles \cite{Fujii:2017ekh,Kumar:2021umc}.

As the lepton colliders offer a significant advantage in terms of QCD background compared to hadron colliders, 
the search for BSM particles in jet enriched signals is much simpler in lepton colliders compared to the LHC.
Additionally, a muon collider \cite{AlAli:2021let} will be advantageous in terms of centre of mass energy reach as well, producing much heavier BSM particles. The proposed reach of muon collider is close to 10 TeV.
There are also proposals for $\gamma \gamma$ collider, which can be constructed out from the future colliders with  
electron-positron\cite{Ellis:2022uxv} or muon -antimuon beams \cite{Klasen:1997zv}.

When searching for the BSM particles, the signature considered is usually based on the assumption that 
only one kind of BSM particle is being produced at the collision and it is decaying directly to the SM particles. However, there are scenarios where more than one  
BSM multiplets exist within a theory, and they interact with each other. 
Large multiplets of fermions and scalars are expected in various theories such as Grand Unified Theory (GUT)\cite{Raby:2006sk}, 
Left-Right Symmetric Model\cite{deAnda:2014dba}, Little Higgs \cite{Kumar:2020yco} and Composite Higgs Models \cite{Panico:2015jxa}.

In this contribution I discuss a minimal scenario based on \cite{Nomura:2017abu} where two types of BSM particles, a fermionic quintuplet and a scalar quartet interact with each other before decaying into SM particles. The decay patterns of these BSM particles 
highly depend on the masses of the BSM particles and the couplings. I will also show that this model yields distinctive 
signatures characterized by high lepton and jet multiplicities, providing a promising opportunity to detect them at future linear collider experiments.

\section{Model}
Models with large multiplets are useful to address the origin of neutrino mass \cite{Kumericki:2012bh}. 
Moreover the neutral fermions act as a dark matter candidates and predict correct relic density \cite{KumarAgarwalla:2018nrn} .
These models are useful to address the tension in muon (g-2) measurement as well \cite{Nomura:2022wck}.
I choose a simple multiparticle scenario with one fermion and one scalar multiplet\cite{Nomura:2017abu}\cite{Kumar:2021umc}, given by $\Sigma= (\Sigma_1^{++},\Sigma_1^{+},\Sigma^{0}, \Sigma_2^{-},\Sigma_2^{--})$ 
and $\Phi= (\phi^{++}, \phi_1^{+}, \phi^0, \phi_2^{-})$ which transforms as (1,5,0) and (1,4,1/2). 
The Gauge interaction terms are
\begin{equation}
\mathcal{L}_{gauge} =\bar \Sigma_R \gamma^\mu i D_\mu \Sigma_R +|D_\mu \Phi|^2. 
\end{equation}
The Yukawa interaction can be expressed as 
\begin{equation}
-\mathcal{L}_{Y}= (y_{\ell})_{ii} \bar L_{L_i} H e_{R_i} +(y_{\nu})_{ij} [\bar L_{L_i} \tilde\Phi \Sigma_{R_j} ]
 +  (M_{R})_i [\bar \Sigma^c_{R_i} \Sigma_{R_i}] + {\rm h.c.}.
 \end{equation}
Depending on the masses of the multiplets, we get two scenarios: $M_\phi> M_\Sigma$ and $M_\phi< M_\Sigma$.
$M_\Sigma > M_\phi$ allows the components of $\Sigma$ to decay to the components of $\Phi$ which is studied in \cite{Kumar:2021umc}. 
Under this assumption $\Phi$ decays to the gauge bosons only (100\% branching ratio), hence the decays of $\Phi$ are purely {\it fermiophobic}.

The scenario with $M_\phi> M_\Sigma$ allows the following decay modes of $\Phi$.
\begin{eqnarray*}
\phi_1^{\pm} &\rightarrow& \Sigma^{0} l^{\pm},  \Sigma^{\pm} \nu, W^{\pm} Z \\
\phi_2^{\pm} &\rightarrow&  \Sigma^{\pm} \nu,  \Sigma^{\pm \pm} l^{\mp}, W^{\pm} Z  \\
\phi^{\pm \pm}  &\rightarrow&  \Sigma^{\pm \pm} \nu, \Sigma^{\pm} l^{\pm},W^{\pm} W^{\pm} \\
\phi^{0} &\rightarrow&  \bar{\nu}_l \Sigma^{0}, \bar{l} \Sigma^{-}, W^{+} W^{-}\\
\end{eqnarray*}
Here, the components of $\Phi$ not only decay directly to the SM gauge bosons, but can also decay to the SM leptons and
components of $\Sigma$.

Now, when $y_{\nu}\sim 1$, we observe that both $\phi_1^{\pm}$ and $\phi_2^{\pm}$ prefers to decay to the fermions but when $y_{\nu}<< 1$, $\phi_1^{\pm}$ prefers to decay to the gauge bosons whereas $\phi_2^{\pm}$ still decays to the fermions in most cases. Hence the decays of $\phi_1^{\pm}$ are both fermiophilic and fermiophobic but the decays of $\phi_2^{\pm}$ are always fermiophilic. The branching ratios are written explicitly in Table\ref{tab:BR_fermiophilic}.
This behaviour leads to unique final states when these particles are produced at collider experiments.
\begin{table}
    \centering
    \begin{tabular}{|c|c|c|}
         \hline
         Decay mode & BR($y_{\nu}\sim 1$) &BR($y_{\nu}<< 1$) \\ 
         \hline
         $\phi_1^{\pm} \rightarrow W^{\pm} Z$ & $<10^{-3}$& 0.9\\
         $\phi_1^{\pm} \rightarrow \Sigma^{0} l^{\pm}$ & $ 0.4$& 0.04 \\
         $\phi_1^{\pm} \rightarrow \Sigma^{\pm} \nu_l$ & $0.6$& 0.06 \\
         \hline
         $\phi_2^{\pm} \rightarrow W^+ Z$ & $<10^{-5}$ & 0.15\\
         $\phi_2^{\pm} \rightarrow \Sigma^+ \nu_l$ & $0.2$ & 0.18\\
         $\phi_2^{\pm} \rightarrow \Sigma^{++} l^{\pm}$ & $0.8$ & 0.67\\
        \hline
    \end{tabular}
    \caption{Branching Ratio of the scalars (500 GeV)for the fermiophilic and fermiophobic scenario .}
    \label{tab:BR_fermiophilic}
\end{table}

When $\Phi$ develops a vev, $v_4 / \sqrt{2}$, we obtain a Dirac mass term connecting the SM neutrinos 
and the neutral component of the quintuplet, which will act as a off diagonal entry in the mass matrix \cite{Nomura:2017abu}. 
This allows the the components of $\Sigma$ decay to the SM leptons via the following decay modes:
\begin{equation}
\Sigma^{0} \rightarrow l^- W^+,~~~  \Sigma^{\pm} \rightarrow \nu W^{\pm} ,~~~\Sigma^{\pm \pm} \rightarrow  W^{\pm} l^{\pm}
\end{equation}
\section{Collider signatures}
\label{sec-1}
\begin{figure*}
\centering
\vspace*{1cm}       
\includegraphics[width=6.2cm,height=5cm]{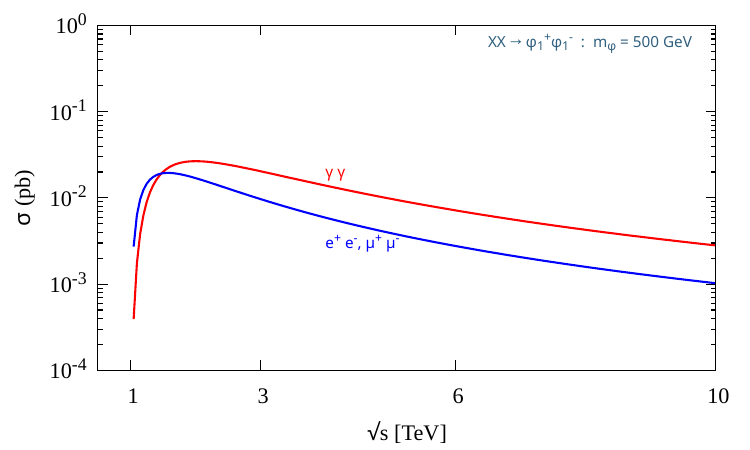}
\includegraphics[width=6.2cm,height=5cm]{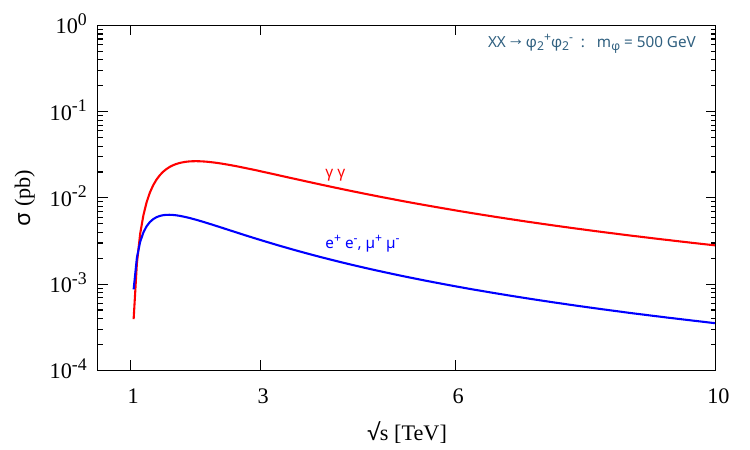}
\caption{Production cross section for $\phi_1^{+} \phi_1^{-}$ (left) and  $\phi_2^{+} \phi_2^{-}$ (Right) at $e^+ e^-$ or $\mu^+ \mu^-$ (blue) and $\gamma \gamma$ (red) collider.}
\label{fig-1}       
\end{figure*}

Pair production of the scalars are simulated using Madgraph \cite{Alwall:2014hca} with ISR effects included.
The pair production cross sections for the two singly charged scalars of mass 500 GeV are shown in figure~\ref{fig-1}, at electron-positron collider and muon collider (blue lines) and at $\gamma\gamma$ collider (red line). 
In this plot, the centre of mass energy varies from 1 TeV to 10 TeV, where 10 TeV is the highest proposed reach of muon collider. Only smaller masses of $\Phi$ are reachable at 500 GeV lepton collider such as ILC. 
Due to coupling with photon, charged scalars can be produced with a considerably larger cross-section at $\gamma\gamma$ colliders. 

Production and decay of the scalars are very interesting due to two different kinds of decay modes as discussed in the earlier section. When $M_{\Phi} > M_{\Sigma}$ and the decay modes are fermiophilic ($y_{\nu}= 1$), some of the final states are unique such as the following: 
\begin{eqnarray*}
\phi_1^{+} \phi_1^{-} & \rightarrow&  \Sigma^{0} l^{+} ~\Sigma^{0} l^{-} \rightarrow  W^+ l^- l^+ W^- l^+ l^- ~~~ (\sigma\times BR\sim ~~1.6 fb)\\
\phi_2^{+} \phi_2^{-} & \rightarrow&  \Sigma^{+ +} l^- ~\Sigma^{- -} l^+  \rightarrow  W^+ l^+ l^- W^- l^- l^+ ~~(\sigma\times BR \sim ~~1.28 fb)\\
\end{eqnarray*}
The $\sigma\times BR$ numbers are calculated for 1 TeV lepton collider and for $m_{\phi}=500$ GeV.
Even though $\phi_1^{\pm}$ and $\phi_2^{\pm}$ show distinct decay patterns, the final state is the same. Hence it is crucial to take contribution from both the processes for the final state with two opposite sign lepton pairs with two oppositely charged 
$W$ bosons or their decay products. There are many possibilities of the final states with multiple leptons and jets and the lepton and the jet multiplicities are 
larger than those considered in the searches by ATLAS and CMS. One example of such final state will be 5 lepton + 2 jets  with two opposite sign lepton pairs. These states will have very small backgrounds at the lepton colliders.

\begin{figure*}
\centering
\vspace*{1cm}       
\includegraphics[width=6.2cm,height=5cm]{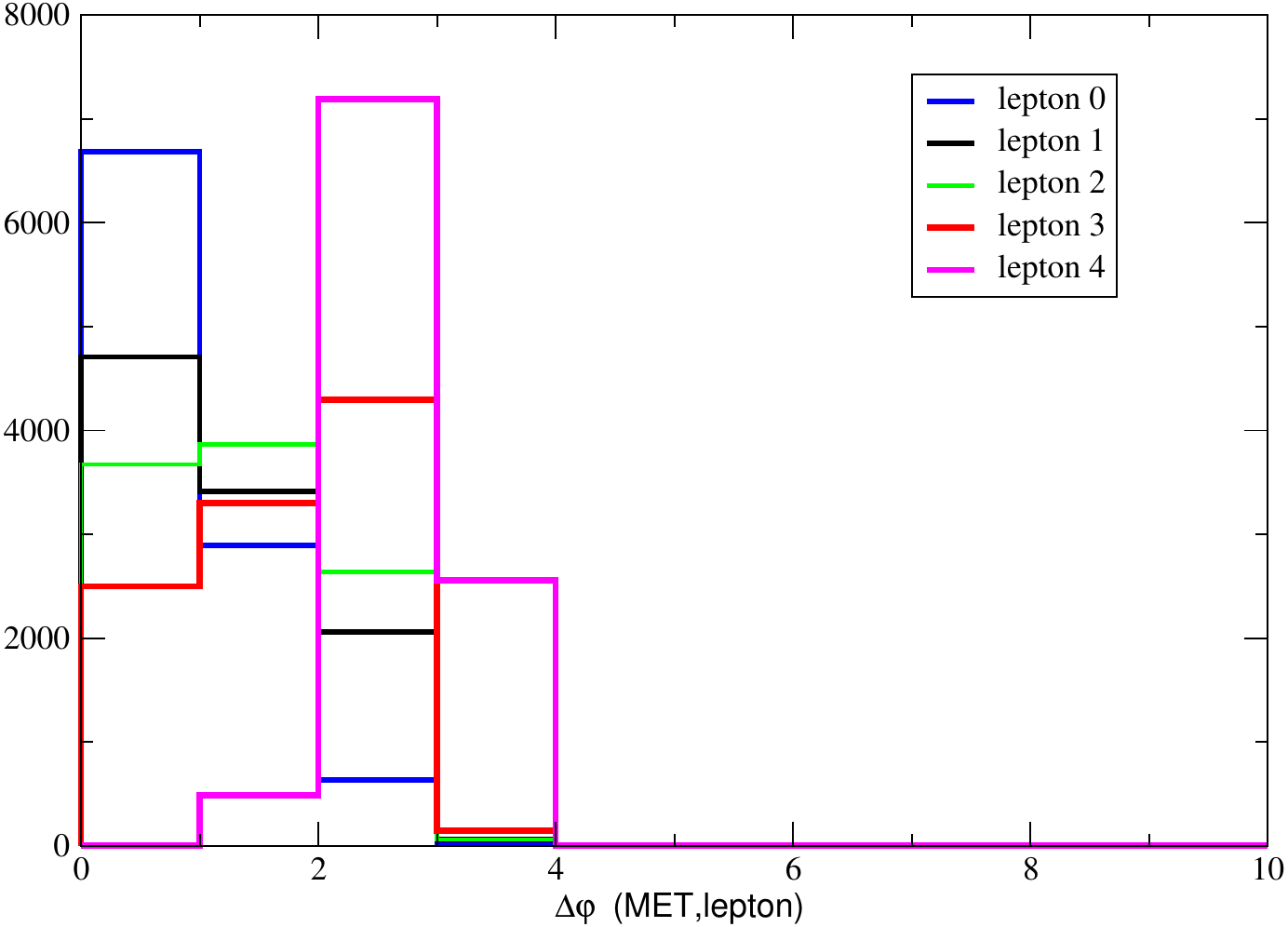}
\includegraphics[width=6.2cm,height=5cm]{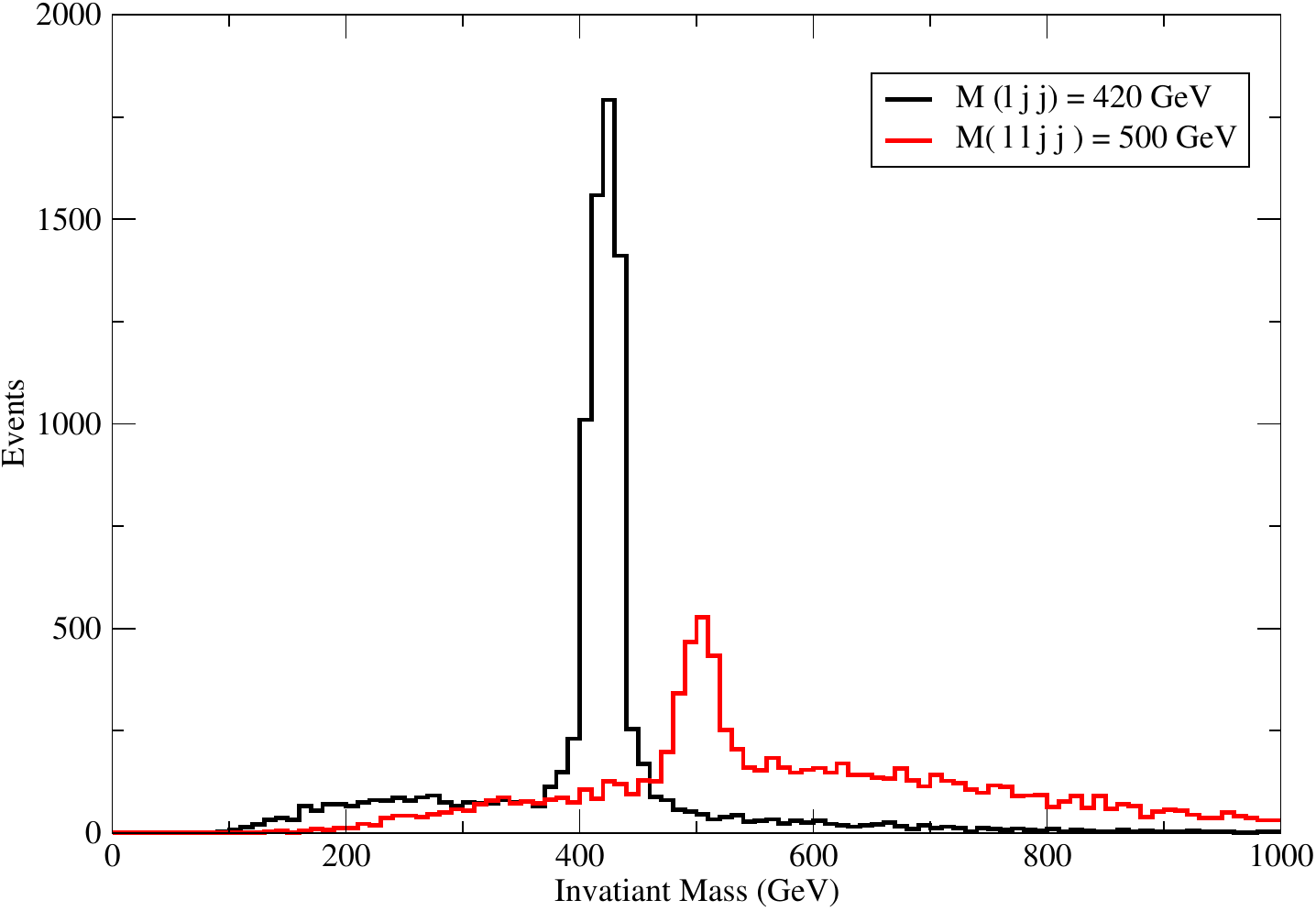}
\caption{Left: ($\Delta \Phi $(MET, lepton) distribution for all leptons in 5 lepton 2 jets channel at lepton collider. Right: Three body and four body invariant mass distribution for $M_{\Sigma}-M_{\Phi}=80$ GeV.}
\label{fig-2}       
\end{figure*}

Signal events with 5 lepton + 2 jets signature, with two opposite sign pairs of leptons are identified by requiring large angular separation  between the lepton and the MET and small separation between the lepton pairs. In figure~\ref{fig-2} (left) we show the the plot for ($\Delta \Phi $(MET, lepton)) for all 5 leptons. The lepton closest to MET is identified as the isolated lepton (lepton 0). (lepton 1, lepton 2) and (lepton3, lepton 4) are the two opposite sign pairs. Jets are selected with di-jet invariant mass close to the $W$ mass. The three body invariant mass of one lepton from the opposite sign lepton pairs along with the two jets are successfully reconstructed, which peaks at $M_{\Sigma}=420 GeV$. The four body invariant mass of one lepton pair with two jets are reconstructed to give $M_{\Phi}=500$GeV, see figure~\ref{fig-2} (right).

For the other scenario, when $y_{\nu}<< 1$, the decay modes of the scalars are fermiophobic. Hence we observe the following unique signatures.
\begin{eqnarray*}
\phi_1^{+} \phi_1^{-} \rightarrow W^+ Z ~W^- Z ~~~(\sigma\times BR\sim ~~8.1 fb)\\
\phi_2^{+} \phi_2^{-}\rightarrow \Sigma^{+ +} l^- ~\Sigma^{- -} l^+ ~~\rightarrow W^+ l^+ ~l^- W^- l^- l^+ ~~(\sigma\times BR\sim ~~0.9 fb)\\
\end{eqnarray*}
Note that here, $\phi_1^{\pm}$ decays to ($W^{\pm},Z$) only whereas $\phi_2^{\pm}$ still decays to the fermions. Hence 
$\phi_2^{\pm}$ will only contribute in the final state of 5 lepton + 2 jets or 4 leptons + 4 jets with two opposite sign pair 
of leptons. 
\section{Conclusion}
A model featuring a fermionic quintuplet and a scalar quartet is discussed. These exotic scalars once produced at the colliders, choose to 
decay to the exotic fermions and then the fermions decay into the SM particles. This decay chain offers unique signatures with high lepton and jet multiplicities. These signatures are very promising as they will be accompanied by very low background at the future lepton colliders.
International Linear Collider and future muon collider experiments offers a clean environment to study these unique 
BSM signals with high lepton and jet multiplicity and the reconstruction of the exotic fermion and scalar mass is also possible. 
Muon colliders, on the other hand, could reach higher energies 
and produce heavier BSM particles. In conclusion, the future lepton colliders offer unique possibilities to search for 
Beyond Standard Model particles and potentially revealing new insights into the fundamental nature of our universe.
%

\end{document}